\documentclass{FBSart}
\usepackage{amsfonts}
\usepackage{amssymb}
\usepackage{graphicx}

\newcommand{\He}{{}^3\mathrm{He}}
\newcommand{\Hh}{{}^3\mathrm{H}}

\title{%Challenges and achievements in %the ab-initio 
Three- and Four-Body Scattering Calculations including the Coulomb Force}

\author{A. Deltuva}

\institute{Centro de F\'{\i}sica Nuclear da Universidade de Lisboa,
P-1649-003 Lisboa, Portugal}

\runningauthor{A.\,Deltuva}
\runningtitle{Three- and Four-Body Scattering Calculations 
including the Coulomb Force}

\begin{document}
\maketitle

\begin{abstract}
The method of screening and renormalization for including the Coulomb 
interaction in the framework of momentum-space integral equations is  
applied to the  three- and four-body nuclear reactions.
The Coulomb effect on the observables and the
ability of the present nuclear potential models to describe the
experimental data is discussed. 
\end{abstract}

%%%%%%%%%%%%%%%%%%%%%%%%%%%%%%%%%%%%%%%%%%%%%%%%%%%%%%%%%%%%%%%%%%%%%%%%%%%%%%%
%\section{Introduction \label{sec:intro}}

The Coulomb interaction, due to its long range, does
not satisfy the mathematical properties required for the formulation
of the standard scattering theory. However, since in nature the Coulomb 
potential is always screened,
one could expect that the physical observables
become insensitive to the screening provided it takes place at sufficiently
large distances $R$ and, therefore, the $R \to \infty$ limit
should correspond to the proper Coulomb. This was proved by
Taylor \cite{taylor:74a} in the context of  the two-particle system:
 though the  on-shell screened Coulomb transition matrix 
 diverges in the $R \to \infty$ limit, after renormalization 
by (an equally) diverging phase factor 
it converges as a distribution to the well known proper Coulomb amplitude
and therefore yields identical results for the physical observables.
A similar renormalization relates screened and proper Coulomb wave functions
\cite{gorshkov:61}.

The method of screening and renormalization  can be used for the systems
with more particles \cite{alt:80a}, albeit with some limitations.
Here we briefly recall the procedure which is described in detail in 
ref.~\cite{deltuva:05a}.
In the  transition operators derived from nuclear plus screened Coulomb 
potentials one has to isolate the diverging screened Coulomb contributions 
in the form of a two-body on-shell transition matrix 
and two-body wave function with known renormalization properties.
This can be achieved using the two-potential formalism
 as long as in the initial/final states
there are no more than two charged bodies (clusters).
At the same time this procedure separates long-range and 
Coulomb-distorted short-range parts of the transition amplitude,
the former being the two-body on-shell transition matrix
derived from the screened  Coulomb potential between
the centers of mass (c.m.) of the two charged bodies 
that is present in the elastic scattering only.
After renormalization this contribution converges  towards its  
$R \to \infty$ limit very slowly but the result, the pure Coulomb amplitude 
of two-body nature, is known analytically. 
The remaining part of the elastic scattering amplitude
as well as the amplitudes for transfer and breakup are 
complicated short-range operators that are externally distorted by Coulomb.
However, due to their short-range nature, convergence with $R$ 
after the renormalization by the corresponding phase factors is fast
and, therefore,  they are calculated numerically
at finite $R$ using the  standard scattering theory and making sure that
$R$ is large enough for the convergence of the results.
We solve  Faddeev-like Alt, Grassberger, and Sandhas (AGS) equations
for three- and four-particle scattering \cite{alt:67a,grassberger:67}
using the momentum-space partial-wave basis as described 
in detail in refs.~\cite{chmielewski:03a,deltuva:03a,deltuva:07a}
for three- and four-nucleon scattering without the Coulomb force.
However, the screened Coulomb interaction,
due to its longer range, compared to the nuclear interaction, 
brings additional difficulties: quasisingular nature of the potential
and slow convergence of the partial-wave expansion.
The right choice of the screening is essential in resolving 
those difficulties. The convergence of the partial-wave expansion with
our new screening function \cite{deltuva:05a} is fast enough and thereby
allows us to avoid the approximations used in the previous implementations 
\cite{alt:94a,alt:02a} of the screening and renormalization approach 
and obtain reliable results.

The most important criterion for the reliability of the screening and
renormalization method is the convergence
of the observables with the screening radius $R$ used to calculate the 
Coulomb-distorted short-range part of the amplitudes.
Numerous examples can be found in 
refs.~\cite{deltuva:05a,deltuva:06b,deltuva:07b}.
In most cases the convergence is impressively fast and only becomes slower
for the observables at very low energies.
Furthermore, as demonstrated in ref.~\cite{deltuva:05b},
our results for $p$-$d$ elastic scattering agree well
 over a wide energy range  with
those of ref.~\cite{kievsky:01a} obtained from the variational solution
of the three-nucleon Schr\"odinger equation in configuration space
with the inclusion of an \emph{unscreened} Coulomb potential
and imposing  the proper Coulomb boundary conditions explicitly.

\begin{figure}[!]
\begin{center}
\includegraphics[scale=0.7]{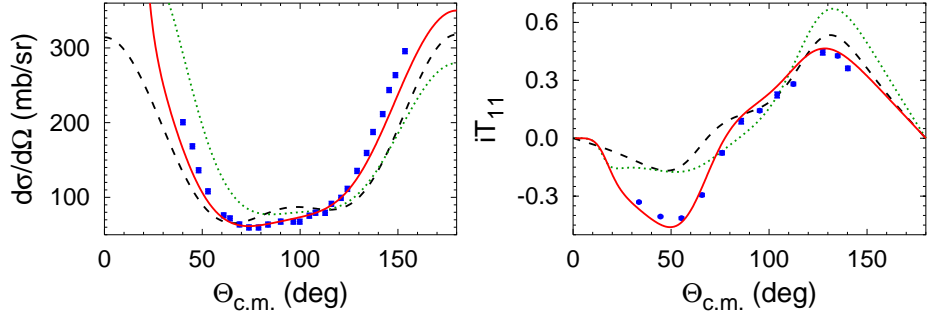}
\end{center} %\vspace{-5mm}
\caption{ \label{f1}
Differential cross section and deuteron vector analyzing power $iT_{11}$
 of $d$-$\alpha$ elastic scattering at $E_d = 4.81$ MeV.
Results derived from the $N$-$\alpha$ potential that is
attractive in S-wave and supports Pauli-forbidden bound state which is 
projected out are shown as solid (dashed) curves with (without) Coulomb.
The results including Coulomb but with  local repulsive $N$-$\alpha$
S-wave potential are given by dotted curves.
The experimental data are from refs.~\cite{bruno:80,gruebler:T} }
\end{figure}

\begin{figure}[!]
\begin{center}
\includegraphics[scale=0.52]{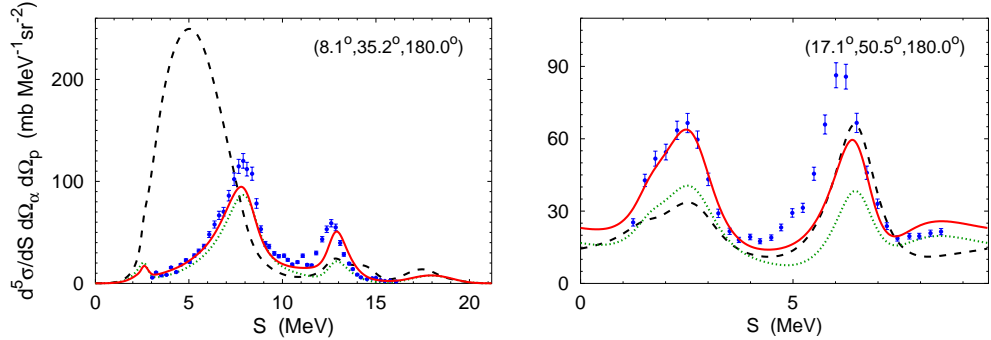}
\end{center} %\vspace{-5mm}
\caption{ \label{f2}
Differential cross section of $\alpha$-$d$ breakup at $E_\alpha = 15$ MeV
in selected kinematical configurations. Curves as in Fig.~\ref{f1}.
The experimental data are from ref.~\cite{bruno:80,gruebler:T} }
\end{figure}

The present method was used to study  three-nucleon hadronic and
electromagnetic (e.m.) reactions in
refs.~\cite{deltuva:05a,kistryn:06a,sauer:fb18,sagara:fb18}.
Furthermore, it was applied to the nuclear reactions
dominated by three-body degrees of freedom like $\alpha + d$ \cite{deltuva:06b},
$d+{}^{12}\mathrm{C}$, and  $p+{}^{11}\mathrm{Be}$ \cite{deltuva:07d,crespo:07a}.
Finally, in refs.~\cite{deltuva:07a,deltuva:07b,deltuva:08a}
all elastic and transfer four-nucleon reactions below
three-body breakup threshold have been studied.
The importance of the Coulomb at low energies is demonstrated in  
Fig.~\ref{f1} for elastic $d$-$\alpha$ scattering. It may be very strong
at all energies in $p$-$d$ breakup and three-body e.m. disintegration of $\He$
in kinematical regimes with low relative $pp$ energy
where the Coulomb repulsion converts the cross section
peak obtained in the absence of Coulomb into a minimum as can be seen
in the the experimental data as well \cite{kistryn:06a,sagara:fb18}.
However, even after the inclusion of the Coulomb interaction
and the three-nucleon force
some  discrepancies between experiment and theory
like the space star anomaly in $p$-$d$ breakup \cite{deltuva:05a,sagara:fb18}
and the $A_y$-puzzle in $p$-$d$ \cite{deltuva:05a,kievsky:01a}
and $p$-$\He$ \cite{deltuva:07b,viviani:01a} elastic
scattering still persist. Furthermore, $A_y$ is described quite well
in the $n$-$\He$ and $p$-$\Hh$ elastic scattering  but
not in the  $p+\Hh \to n + \He$ transfer reaction.
A very strong Coulomb effect manifests itself in the $\alpha$-$d$ breakup
where the shift of $\alpha p$ $P$-wave resonance
position leads to the corresponding shifts of the differential cross
section peaks as shown in Fig.~\ref{f2}.
In addition, Figs.~\ref{f1} and \ref{f2} as well as the results of
ref.~\cite{deltuva:06b} demonstrate the superiority
of the attractive $N$-$\alpha$ S-wave potentials supporting a Pauli-forbidden
bound state that is projected out over the local repulsive S-wave potentials
which, because of their simplicity, are very often used in the
configuration space calculations of resonances and e.m. reactions.

In conclusion, the Coulomb interaction between the charged
particles was included in few-body scattering calculations
using the old idea of screening and renormalization \cite{taylor:74a}
but with novel practical realization 
that avoids all the approximations of the previous works \cite{alt:94a,alt:02a}
and yields fully converged results.

\begin{acknowledge}
This work has been performed in collaboration with A.~C.~Fonseca and 
P.~U.~Sauer. The author is supported by the Funda\c{c}\~{a}o para a 
Ci\^{e}ncia e a Tecnologia (FCT) grant SFRH/BPD/34628/2007.
\end{acknowledge}

%%%%%%%%%%%%%%%%%%%%%%%%%%%%%%%%%%%%%%%%%%%%%%%%%%%%%%%%%%%%%%%%%%%%%%%%%%%%%
%\bibliographystyle{fewbody}
%\bibliography{abbrev,pre80,80-89,90-99,200x,clmb,ad,4N,hann,book,numerics,exp}

\begin{thebibliography}{10}

\bibitem{taylor:74a}
Taylor, J.~R.: Nuovo Cim. {\bf B23},  313  (1974)

\bibitem{gorshkov:61}
Gorshkov, V.~G.: Sov.~Phys.-JETP {\bf 13},  1037  (1961)

\bibitem{alt:80a}
Alt, E.~O., Sandhas, W.: Phys.~Rev. {\bf C21},  1733  (1980)

\bibitem{deltuva:05a}
Deltuva, A., Fonseca, A.~C., Sauer, P.~U.: 
Phys.~Rev. {\bf C71}, 054005  (2005);  Phys.~Rev.~Lett. {\bf 95},  092301;
Phys.~Rev. {\bf C72}, 054004  (2005); 
Phys.~Rev. {\bf C73}, 057001  (2006); 
Annu.~Rev.~Nucl.~Sci. {\bf 58},  27  (2008)

\bibitem{alt:67a}
Alt, E.~O., Grassberger, P., Sandhas, W.: Nucl.~Phys. {\bf B2},  167  (1967)

\bibitem{grassberger:67}
Grassberger, P., Sandhas, W.: Nucl. Phys. {\bf B2},  181  (1967); 
Alt, E.~O., Grassberger, P., Sandhas, W.: JINR report No. E4-6688 (1972)

\bibitem{chmielewski:03a}
Chmielewski, K., et al.:  Phys.~Rev. {\bf C67},  014002  (2003)

\bibitem{deltuva:03a}
Deltuva, A., Chmielewski, K., Sauer, P.~U.: Phys.~Rev. {\bf C67}, 034001 (2003)

\bibitem{deltuva:07a}
Deltuva, A., Fonseca, A.~C.: Phys.~Rev. {\bf C75},  014005  (2007)

\bibitem{alt:94a}
Alt, E.~O., Rauh, M.: Few-Body Systems {\bf 17},  121  (1994)

\bibitem{alt:02a}
Alt, E.~O., et al.:  Phys.~Rev. {\bf C65},  064613  (2002)

\bibitem{deltuva:06b}
Deltuva, A.: Phys.~Rev. {\bf C74},  064001  (2006)

\bibitem{deltuva:07b}
Deltuva, A., Fonseca, A.~C.: Phys.~Rev.~Lett. {\bf 98},  162502  (2007);
Phys.~Rev. {\bf C76},  021001  (2007)

\bibitem{deltuva:05b}
Deltuva, A., et al.: Phys.~Rev. {\bf C71},  064003  (2005)

\bibitem{kievsky:01a}
Kievsky, A., Viviani, M., Rosati, S.: Phys.~Rev. {\bf C64}, 024002 (2001)

\bibitem{kistryn:06a}
Kistryn, S.,  et al.: Phys.~Lett. {\bf B641},  23  (2006)

\bibitem{sauer:fb18}
Deltuva, A., Fonseca, A.~C., Sauer, P.~U.: Nucl.~Phys. {\bf A790}, 344c (2007)

\bibitem{sagara:fb18}
Sagara, K., et al.: Nucl.~Phys. {\bf A790},  348c  (2007)

\bibitem{deltuva:07d}
Deltuva, A., et al.: Phys.~Rev.  {\bf C76},  064602  (2007)

\bibitem{crespo:07a}
Crespo, R., et~al.: Phys.~Rev. {\bf C76},  014620  (2007)

\bibitem{deltuva:08a}
Deltuva, A., Fonseca, A.~C., Sauer, P.~U.: Phys.~Lett. {\bf B660}, 471 (2008)

\bibitem{bruno:80}
Bruno, M., et al.: Lett. Nuovo Cim. {\bf 27},  265  (1980)

\bibitem{gruebler:T}
Gr\"uebler, W., et al.: Nucl. Phys.{\bf A134},  686  (1969)
%; {\bf A148},  380  (1970); {\bf  A148},  391  (1970).

\bibitem{viviani:01a}
Viviani, M., et al.: Phys.~Rev. Lett. {\bf 86}, 3739  (2001)

\bibitem{koersner:77}
Koersner, I., et al.:  Nucl. Phys. {\bf A286},  431  (1977)

\end{thebibliography}

\end{document}